\documentclass[%
 reprint,
 amsmath,amssymb,
 aps, dvipsnames,
]{revtex4-2}

\usepackage{graphicx}
\usepackage{dcolumn}
\usepackage{bm}

\usepackage{siunitx}
\usepackage{braket}
\usepackage{natbib}
\usepackage{enumitem}
\usepackage{color}
\usepackage[hidelinks]{hyperref}
\newcommand{\mycomment}[1]{}

\begin{document}

\preprint{APS/123-QED}

\title{Phase structure of below-threshold harmonics in aligned molecules:\\ a few-level model system}

\author{Samuel Sch\"opa$^1$}
\author{Lina Bielke$^1$}
\author{Falk-Erik Wiechmann$^{1,2}$}
\author{Franziska Fennel$^{1,2}$}
\author{Dieter Bauer$^{1,2}$}

\affiliation{%
$^1$Institute of Physics, University of Rostock, 18051 Rostock, Germany.\\
$^2$Department Life, Light \& Matter, University of Rostock, 18051 Rostock, Germany.
}
\date{\today}

\begin{abstract}
We utilize few-level model systems to analyze the polarization and phase properties of below-threshold harmonics generated from aligned molecules. In a two-level system, we find that the phase of emitted harmonics undergoes a distinct change. For harmonics with photon energies below the transition energy between the dominant field-dressed states, the phase alternates by $\pi$ between successive odd harmonic orders. In contrast, the phase remains constant for harmonics above the transition energy. Exploiting this behavior, we construct a four-level model composed of two uncoupled two-level systems aligned along orthogonal directions. We demonstrate that with selected transition frequencies lower-order harmonics follow the polarization of the linearly polarized driving field while higher-order harmonics exhibit a mirrored polarization.
The model predicts that aligned systems with orthogonal transition dipoles may show analogous phase and polarization features in the below-threshold regime.
\end{abstract}

\maketitle


The mechanisms underlying below-threshold harmonics (BTH) in high-harmonic generation (HHG) from molecules driven by short, intense laser pulses differ fundamentally from the mechanisms responsible for the generation of higher-order above-threshold  harmonics. In the gas phase, the latter typically involves a single active electron undergoing tunnel ionization to the continuum, followed by quasi-free propagation and recombination with the parent ion \cite{Kulander1993, corkum_plasma_1993}. In contrast, below-threshold, non-perturbative harmonics may have substantial contributions produced by bound electrons that propagate under the force of the laser field along different electronic excited states. The time-dependent populations of the electronic states in the molecule and the related time-dependent dipole leads to the emission of harmonic radiation with photon energies below the ionization energy. Compared to above-threshold harmonic orders, the features of the BTH in molecules are much more target sensitive, as they are strongly influenced by the energy levels and the dipole transition matrix elements of the molecules. 
Its sensitivity  makes BTH a valuable tool for high-harmonic spectroscopy \cite{Soifer_2010, Dong_15, Xiong_2017, Peng_20, long2023below} of molecules while their potential for VUV frequency comb generation \cite{yost2009vacuum, Zhu_2021, Xiong_2017} has further expanded their relevance. As a result, interest on BTH has grown significantly in recent years.
The potential for spectroscopic insight comes to the price of complex analysis to unravel the multiple contributions \cite{Xiong_2017}.\\
\indent While most studies apply semi-classical analyses to model BTH \cite{Xiong_2017, hostetter2010semiclassical}, which work well for very strong fields ($I_0\approx10^{14}$~W/cm$^2$), these models are inaccurate for smaller intensities and obscure the role of discrete electronic transitions and their transition directions for the polarization properties of the emitted harmonics. 
On the other hand, a two-level model system (TLS) has been shown to reproduce features of the molecular harmonic response of H$_2^+$ in the long-wavelength regime for more moderate fields \cite{zuo1993harmonic}.
As we are interested in the contribution of the bound-bound electronic transitions to the BTH, we turn to few-level model systems where specific transitions, dipole orientations, and resonances can be precisely controlled. Such simplified models offer a framework to investigate the mechanisms responsible for the complex polarizations of BTH from aligned molecules.
Despite the simplicity of few-level models, they cannot be solved analytically beyond certain edge cases, like the rotating-wave approximation or approximate treatments using Floquet-Green formalisms \cite{martinez2005high}. However, even analytic approaches have revealed highly nontrivial harmonic phase behavior. For example, emitted plateau harmonics exhibit abrupt $\pi$ phase jumps at specific Rabi frequencies \cite{gauthey1997phase}.
These investigations were carried out in a parameter regime with extremely high field strengths where most population transfers between the adiabatic states occur at very distinct times. In such cases analogies to the three-step model can be drawn \cite{faria_two_level_2002}.\\
\indent To make the emission from a TLS similar to the contribution of a bound-bound transition to the BTH in a molecule, we  will investigate the dependence of the harmonic phase in a TLS at more moderate field-strengths and long wavelengths. Building on this approach, we present a model composed of two uncoupled TLS with orthogonal transition dipoles, to investigate how the phase properties of TLS manifest in the harmonic response of molecules with two distinct axes.
\\


Consider the harmonic response of a TLS with the field-free Hamiltonian
\begin{equation}
    \hat{H}_0 = \begin{pmatrix}
-\omega_{10}/2 & 0\\
0 & \omega_{10}/2
\end{pmatrix}. 
\end{equation}
We use atomic units throughout this paper and set the transition frequency to $\omega_{10} = 0.1$. We only allow for a single transition in the $x$-direction, which simplifies the interaction part to 
\begin{equation}
 \hat{H}_1(t) = \begin{pmatrix}
0 & E_x (t) d\\
E_x (t) d & 0
\end{pmatrix},
\end{equation}
with a transition dipole moment of $d = -1.5$  and an electric field of  
\begin{equation}
    E_x(t) = E_0~\text{env}(t) \cos (\omega_L t),
\end{equation} with the laser frequency $\omega_L$ and the Gaussian envelope
\begin{equation}
    \text{env}(t) = e^{-\frac{1}{2\sigma^2}\left(t - \frac{t_{\text{end}}}{2}\right)^2},
\end{equation}
where $t_{\text{end}} = 120 \pi/\omega_L$ and $\sigma = t_{\text{end}}/16$. The electric field is smoothly brought to zero at the edges using a Tukey window.
The total Hamiltonian reads $\hat{H}(t) = \hat{H}_0 + \hat{H}_1(t)$.
To investigate the electron dynamics, we solve the time-dependent Schrödinger equation (TDSE)
\begin{equation}
    i\frac{d}{dt}\textbf{C}(t) = (\hat{H}_0 + \hat{H}_1(t)) \textbf{C}(t) \label{eq_prop}
\end{equation}
for the vector $\textbf{C}(t)$ describing the populations in the basis of the field-free orbitals, i.e.,  $\ket{\Psi (t)} = C_0(t) \ket{\Psi_0 (0)} + C_1(t) \ket{\Psi_1 (0)}$.   The TDSE is solved numerically using an explicit Runge-Kutta method of order 5(4) to obtain the time-dependent dipole $D(t) = d (C_0^*(t) C_1(t) + C_1^*(t) C_0(t))$. The harmonic spectrum is calculated via the  Fourier-transform of the dipole acceleration
\begin{equation}
    H(\omega) = ||\mathcal{F} \{ \ddot{D}(t)\cdot \text{hann}(t) \}||^2, 
    \label{eq_HHG}
\end{equation}
with a hanning window $\text{hann}(t)$.\\

\begin{figure}
    \centering
    \includegraphics[width=\linewidth]{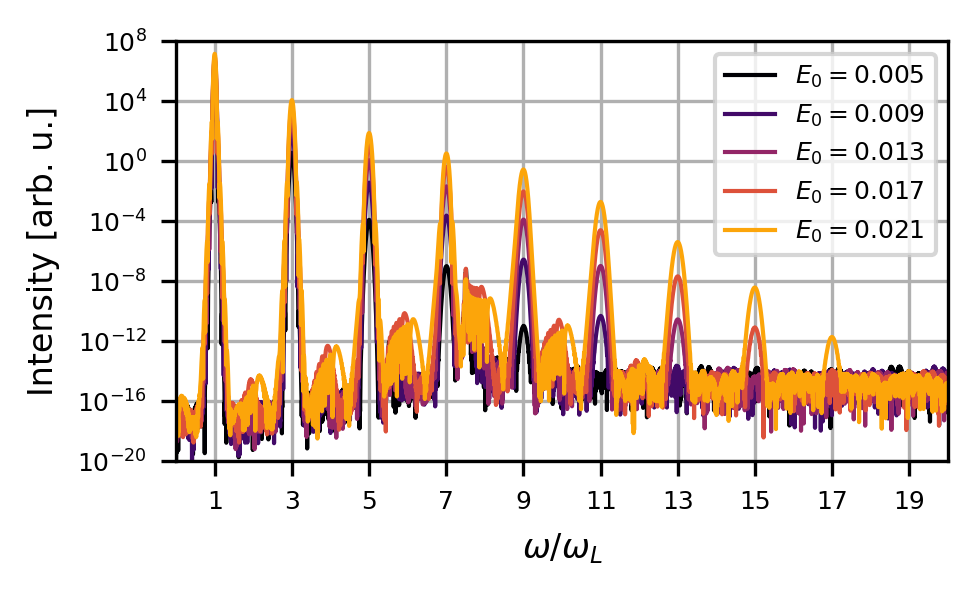}
    \caption{High-harmonic spectrum of a TLS for different field-strengths.}
    \label{fig:Ex_scan}
\end{figure}

Our goal is to study the phase of the harmonics that originate from a single bound–bound transition and its contribution to a molecule’s harmonic spectrum. To this end, we use a TLS and choose the laser parameters such that we circumvent plateau harmonics. This choice is necessary because, in physical systems, very strong fields and short wavelengths would lead to continuum transitions that dominate the harmonic response, which cannot be described within a TLS. Figure~\ref{fig:Ex_scan} shows harmonic spectra for different field strengths $E_0$ at a driving frequency $\omega_L$ that is 7.5 times smaller than the transition energy  $\omega_{10}$ ($\omega_L=\omega_{10}/7.5$).
We observe odd harmonics from order three to 17 as well as non-harmonic spectral features in between. Furthermore, we investigated the phase of the individual harmonics, which is found to be independent of the field strength in our parameter regime. This is in contrast to observations from Gauthey {\em et al.} with much stronger fields, for which avoided crossings of Floquet states lead to phase jumps~\cite{gauthey1997phase}. Our field strengths are too weak to significantly change the Floquet states such that the phase jumps are absent. \\
\begin{figure}
    \centering
    \includegraphics[width=\linewidth]{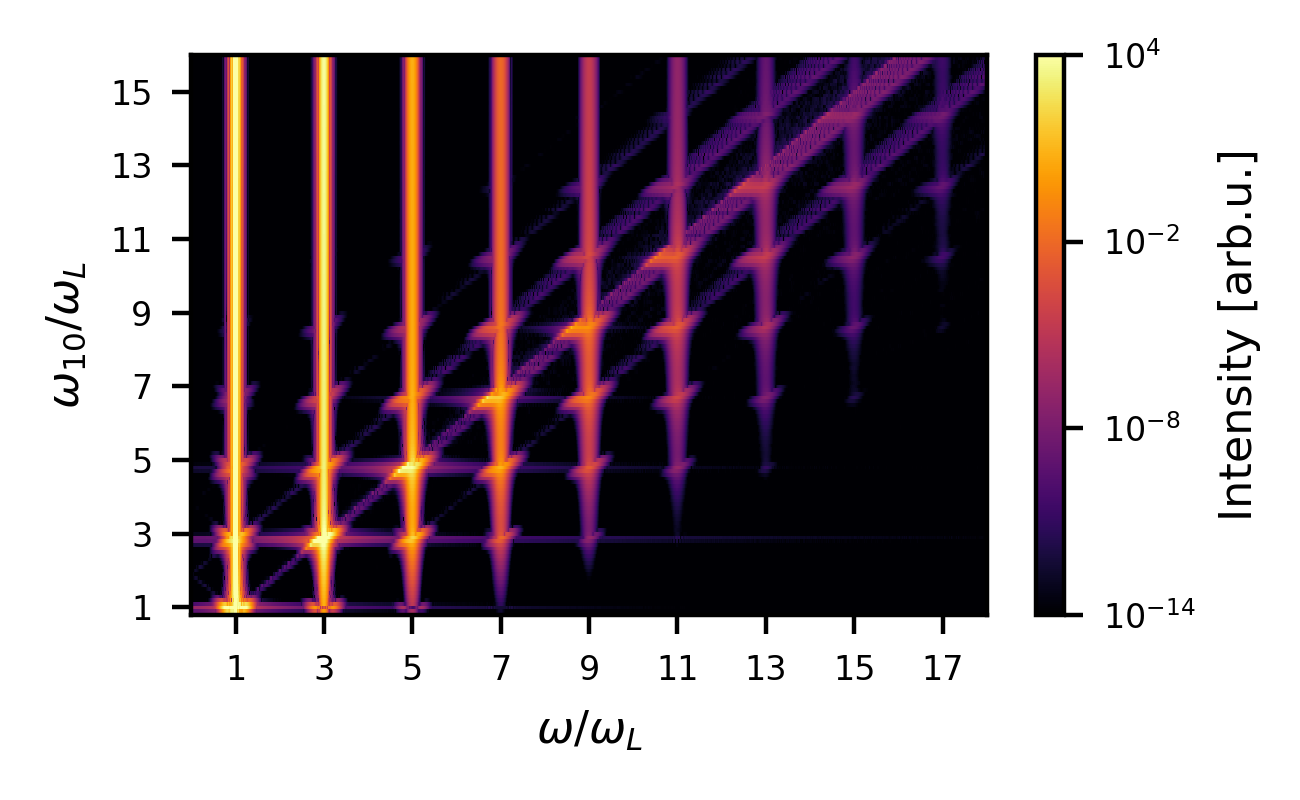}
    \caption{Spectral intensity as a function of the ratio between transition frequency $\omega_{10}$ and laser frequency $\omega_L$ and harmonic order $\omega/\omega_L$. The plotted intensity is capped to the range $\left[10^{-14}, 10^{4}\right]$.}
    \label{fig:2lev_yield}
\end{figure}
\indent However, we find a dependence of the harmonic phase on the frequency of the driving field. In order to provide context for the phase analysis, we calculate the harmonic yield  for a constant field strength $E_0 = 0.015$ and vary the laser frequency $\omega_{L}$.
In Fig.~\ref{fig:2lev_yield} we show the spectral intensity plotted as a function of the ratio $\omega_{10}/\omega_L$ and the harmonic order.
We observe harmonics up to order 17. These harmonics appear as vertical lines. As $\omega_{10}/\omega_L$ increases, higher-order harmonics are progressively enhanced, consistent with \cite{gaarde2001enhancement}. Whenever $\omega_{10}/\omega_L$ is close to an odd integer (with an energy shift due to field-dressing), the entire spectrum is enhanced. Such a global enhancement due to multiphoton resonances has been analyzed theoretically and experimentally in~\cite{toma_1999, gaarde2001enhancement, ganeev2006strong, son2009floquet, ivanov2008resonant}. \\
\indent At $\omega_{10}/\omega_L = 1$, we see the tiny Mollow-triplet of a TLS driven in resonance as side-peaks around the fundamental that are split apart by the Rabi-frequency $\Omega = d E_{0}$.  
For small deviations from  $\omega_{10}/\omega_L = 1$, these satellite peaks move diagonally according to the change in the generalized Rabi frequency $\sqrt{\Omega^2 + (\omega_L-\omega_{10})^2}$, as predicted by the rotating wave approximation. 
Outside strict resonance, additional non-harmonic spectral features appear between the odd harmonics due to superpositions of Floquet states. The primary resonance on the diagonal has additional satellite peaks spaced by integer multiples of $2\omega_{L}$, caused by the periodicity of the Floquet quasi-energy zones \cite{chu2004beyond, hazanov2025high}.\\
\begin{figure}
    \centering
    \includegraphics[width=\linewidth]{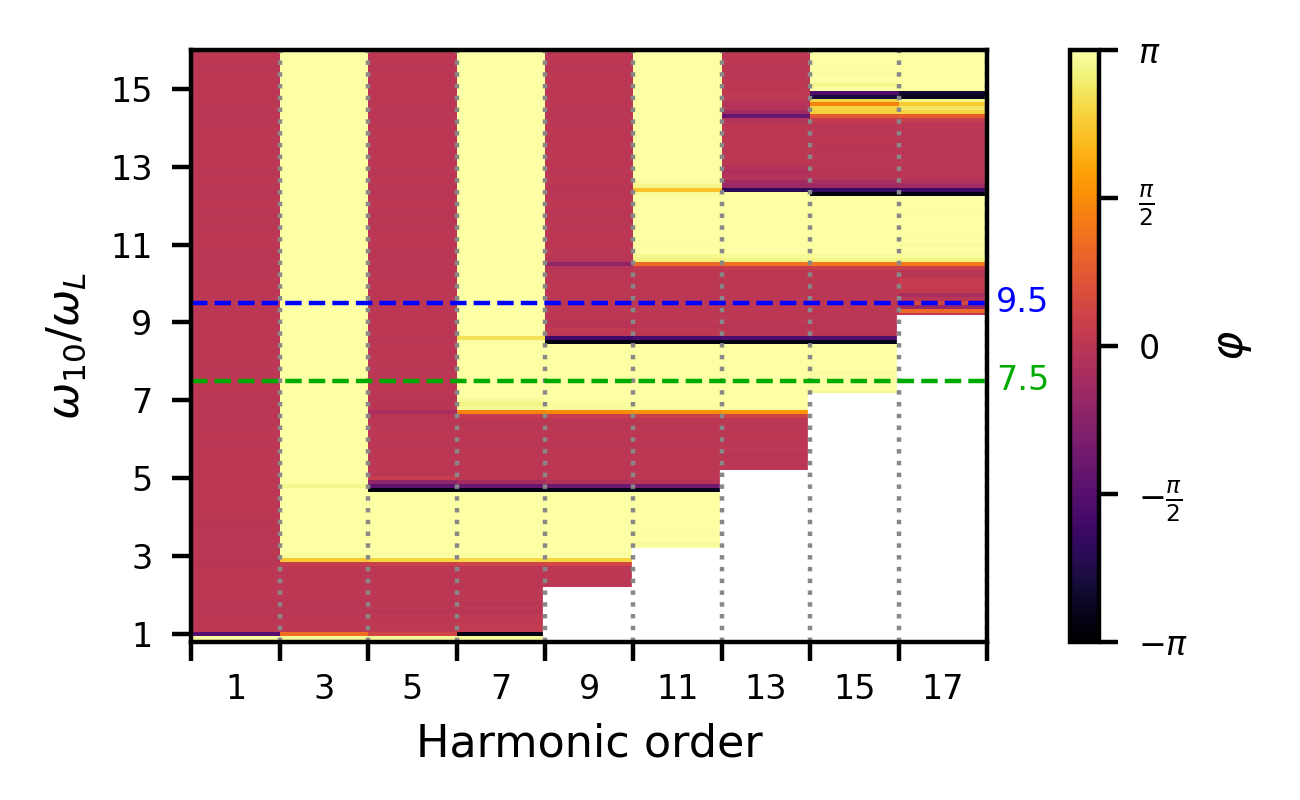}
    \caption{The phase of the different harmonics plotted against the ratio between transition energy and laser frequency. The dashed lines indicate $\omega_{10}/\omega_L = 7.5$ (green) and $\omega_{10}/\omega_L=9.5$ (blue), which are the relevant frequency ratios for the model in Fig.~\ref{fig:HHG_intensity_model} and \ref{fig:HHG_phase_model}.}
    \label{fig:2lev_phase}
\end{figure}
\indent In Fig.~\ref{fig:2lev_phase}, we analyze the phase of the  harmonic orders as a function of $\omega_{10}/\omega_L$. Wherever the system emits odd harmonics, we observe regions with either a phase of zero (red) or $\pi$ (yellow) with respect to the phase of the driving field. Below the  Floquet multi-photon resonance ($\text{harmonic order}<\omega_{10}/\omega_{\rm{L}}$), the phase of the different harmonic orders alternates between $0$ and $\pi$,  whereas above resonance, the phase remains constant. Whenever an odd number of photons matches the gap between the field-dressed states, a crossing appears in the Floquet band structure, altering the phase behavior. A similar observation was made by Gauthey {\em et al.} where a change in the Rabi frequency causes jumps of the harmonic phase precisely at locations of avoided crossings of Floquet states \cite{gauthey1997phase}.  \\
\indent To analyze why the position of the resonance in the spectrum affects the phase of the emitted harmonics, we investigate the origin of the phase with an analytic expression.
Within the adiabatic approximation it is possible to derive a continued-fraction expression for the harmonics emitted by driving a TLS with a periodic laser field \cite{Plaja1993}. Further assuming that each successive harmonic order is substantially weaker than the previous one, the Fourier-transformed Dipole $\widetilde{D}(\omega)$ of the $n$-th harmonic can be expressed as a function of the preceding odd harmonic $n-2$ by 
\begin{equation}
    \widetilde{D}(n\omega_L) = \frac{-e^{-2i\varphi_L} \frac{n-2}{n-1} }{\frac{2 n^2}{n^2 -1} + \frac{\omega_{10}^2 - n^2 w_L^2}{E_0^2 d^2} } \widetilde{D}[(n-2)\omega_L]  \label{eq_pert}
\end{equation}
for the electric field $E(t) = E_0 \cos(\omega_L t - \varphi_L)$ with the phase $\varphi_L$. For a derivation of (\ref{eq_pert}) see \cite{Plaja1993} or the Supplemental Material \cite{SM} (see also Refs. \cite{Plaja1993, Bauer2013} therein). We can extract the phase of the $n$-th harmonic with $\varphi_n = \text{arg}(\widetilde{D}(n\omega_L))$. Note that the only terms in (\ref{eq_pert}) that can be complex are $\widetilde{D}((n-2)\omega_L)$ and $e^{-2i\varphi_L}$. Therefore, within the limits of the approximations, for a real fundamental and $\varphi_L =0$, i.e. a cosine pulse, all higher orders will be real as well. Consequently, the phase of the harmonics calculated with (\ref{eq_pert}) can only assume $\varphi_n \in \{0, \pi\}$. If the prefactor in front of $\widetilde{D}((n-2)\omega_L)$ is negative, the harmonic phase will flip by $\pi$ with each successive harmonic, and if the prefactor is positive, the phase will stay the same. The only term that, for $\varphi_L=0$, can change the sign is the $(\omega_{10}^2 - n^2 w_L^2)/E_0^2 d^2$ term in the denominator. For our parameters, outside strict resonance, it will always be larger than $2 n^2/(n^2 -1)$, as $E_0$ is very small. This gives us the origin of the different behavior above and below the resonance: Below resonance $\omega_{10}^2 - n^2 w_L^2 > 0$, retaining the minus sign in the prefactor, while above resonance, $\omega_{10}^2 - n^2 w_L^2$ is smaller than $0$ which flips the sign in the prefactor. This behavior agrees with the numerical calculations shown in Fig.~\ref{fig:2lev_phase}, except that if a harmonic is driven in resonance, the phase can assume values of $\pm \pi/2$, which does not occur in the analytic expression (\ref{eq_pert}) because the adiabatic approximation breaks down for resonant driving.
Note that the accumulation of the $e^{-2i\varphi_L}$ phase for each successive odd harmonic is caused by the behavior of Fourier-transformations with respect to time-shifts, as a time shift by $\Delta t = \varphi_L/\omega$ will give the $n$-th harmonic the phase factor $e^{-n i \varphi_L}$.
As a result, the phase of the harmonics relative to the phase of the driving field in Fig.~\ref{fig:2lev_phase} is dependent on the carrier-envelope phase (CEP). However, the CEP will not alter the characteristic switch in behavior above the resonance.\\


We now address how such phase behavior could manifest in physical systems. One possible signature is the polarization of harmonics emitted by two-dimensional molecules. Organic molecules can feature two dominant excited states with orthogonal transition dipole moments, such as acenes \cite{Craciunescu2023} and perylene bismides \cite{Wuerthner2004}. In such targets, if the harmonics' emission would be governed primarily by two isolated bound-bound transitions, the response may resemble that of a two-dimensional model composed of two uncoupled TLS with different transition frequencies and orthogonal transition dipoles. In the following, we demonstrate how laser-frequency dependent phase jumps directly translate into distinct polarization patterns of the emitted harmonics. 
We model the 2D system with the field-free Hamiltonian
\begin{equation}
    \hat{H}_0 = \begin{pmatrix}
-\omega_{30y}/2 & 0 & 0 &0\\
0 &-\omega_{21x}/2 & 0&0\\
0&0 & \omega_{21x}/2 &0\\
0&0 & 0& \omega_{30y}/2
\end{pmatrix}.
\end{equation}
Both TLS have the same transition dipole strength of $d=-1.5$ but couple to different electric field components, leading to the interaction part
\begin{equation}
    \hat{H}_1(t) = \begin{pmatrix}
0 & 0 & 0 & E_y (t) d\\
0 & 0 & E_x (t) d & 0\\
0 & E_x(t) d & 0 & 0\\
E_y (t) d & 0 & 0 & 0
\end{pmatrix} 
\end{equation} of the Hamiltonian $\hat{H} = \hat{H}_0 + \hat{H}_1(t)$, with the electric field 
\begin{equation}
    \textbf{E}(t) = \begin{pmatrix}
         E_x (t)\\ E_y (t)
    \end{pmatrix} = 0.015~\text{env}(t) \cos (\omega_L t) \begin{pmatrix}
         \cos \alpha\\ \sin \alpha
    \end{pmatrix},
\end{equation} where $\alpha$ is  the polarization angle relative to the $x$-axis.
The important difference between the two subsystems is their transition frequency of  $\omega_{21x} = 0.1$ and $\omega_{30y} = 0.12\overline{6}$. 
The frequency of the electric field is $\omega_L = 0.1/7.5$, placing the $x$-resonance at $\omega/\omega_L = 7.5$ and the $y$-resonance at $\omega/\omega_L = 9.5$ in the HHG spectrum. 
We solve (\ref{eq_prop}) numerically with the initial condition $\textbf{C}(0) = (1,1,0,0)^\top$. The two-dimensional dipole is obtained by calculating 
\begin{equation}
    \textbf{D}(t) = d \begin{pmatrix}
    C_1^*(t) C_2(t) + C_2^*(t) C_1(t)\\
    C_0^*(t) C_3(t) + C_3^*(t) C_0(t)
    \end{pmatrix},
\end{equation} and the total harmonic intensity with (\ref{eq_HHG}). \\

\begin{figure}
    \centering
    \includegraphics[width=\linewidth]{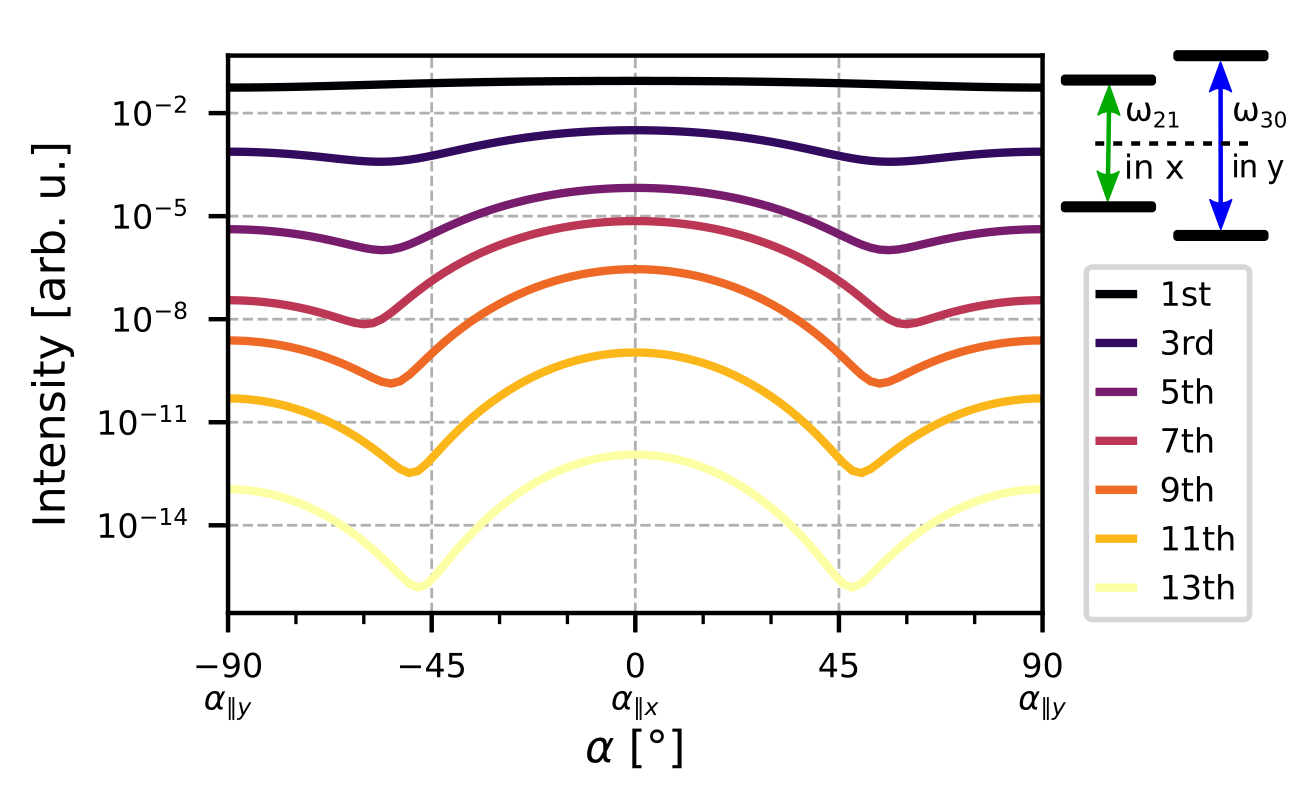}
    \caption{Intensity of the harmonics in the four-level model system with two decoupled two level systems as a function of the polarization angle of the driving field.}
    \label{fig:HHG_intensity_model}
\end{figure}

\begin{figure}
    \centering
    \includegraphics[width=\linewidth]{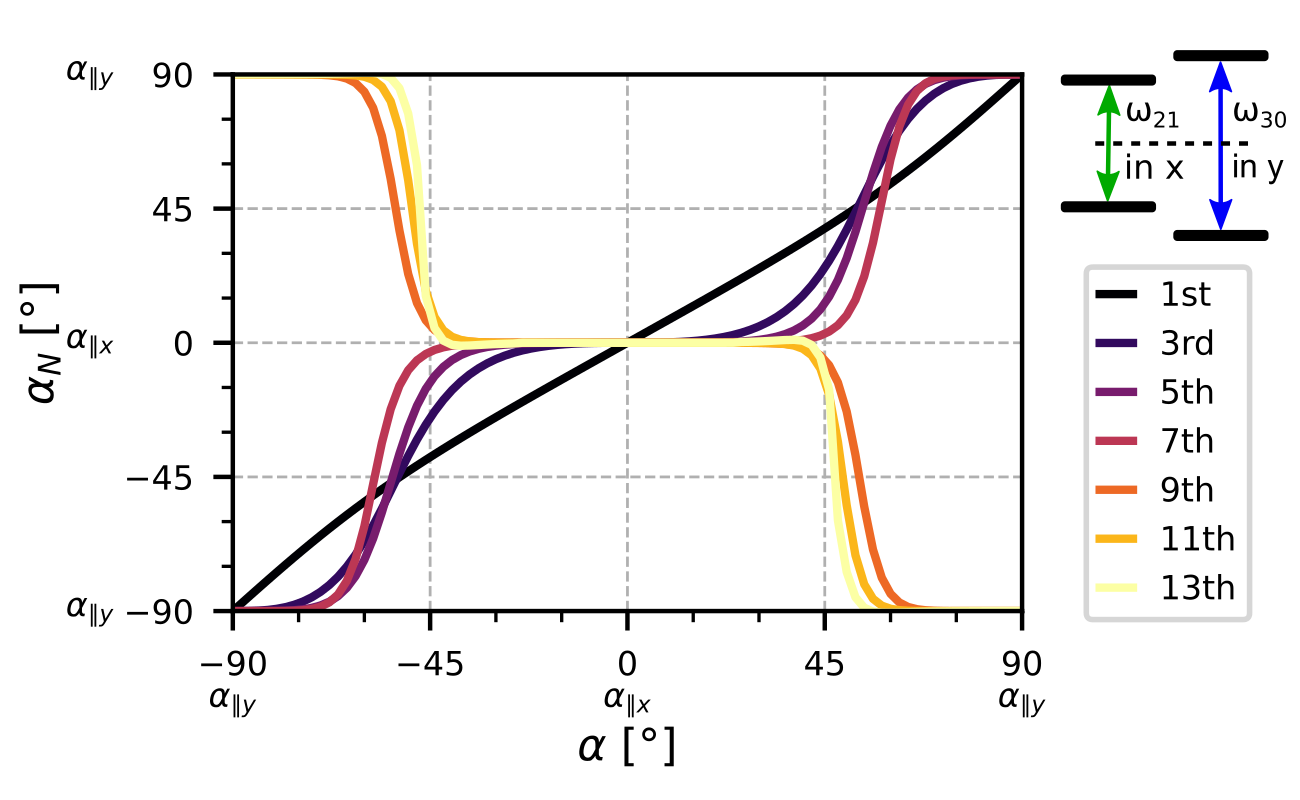}
    \caption{Polarization of the harmonics in the four-level model system with two decoupled two level systems as a function of the polarization of the driving field.}
    \label{fig:HHG_phase_model}
\end{figure}

The phase of the different harmonic orders for the two TLS is marked  with the green~(x-TLS) and blue~(y-TLS) vertical dashed lines in Fig.~\ref{fig:2lev_phase}. For harmonics below the transition energy of the green subsystem ($\text{harmonic order}<\omega_{21}/w_L$, up to order 7), the phases of the emitted harmonics from both subsystems are equal. As the blue subsystem has a transition energy that is $2\omega_L$ larger, the above described change from the alternating phase (below resonance) to constant phase (above resonance) occurs for it at one odd harmonic order higher. Therefore, all harmonics above the seventh of the two subsystems will have a phase shift of $\pi$ relative to each other.\\
\indent Before analyzing the polarization properties of the harmonics, we briefly look at the intensities of the emitted harmonics, which are shown in 
Fig.~\ref{fig:HHG_intensity_model} as a function of the polarization~$\alpha$ of the incoming laser pulse. The intensity shows maxima for an electric field polarized in $x$ and $y$ direction. In between, pronounced minima emerge near intermediate orientations.
The excitation is most efficient when the driving field aligns with the transition dipole moment of a subsystem. Between these orientations, due to the nonlinear scaling of each subsystem's emission with the projection of the electric field on the transition direction, minima in intensity arise. The exact position of the minima also depend on the electronic structure parameters of the two TLS.\\
\indent The polarization of the emitted harmonics is shown in Fig.~\ref{fig:HHG_phase_model} as a function of the polarization~$\alpha$.
Due to the discrete phase shift of $0$ or $\pi$ between the $x$- and the $y$-component, all harmonics are linearly polarized at an angle of  $\alpha_N$ to the $x$-axis. 
The polarization angles of the harmonics 1 to 7 shows a monotonic increase as the polarization of the driving electric field is increased. 
However, due to the different scaling of the harmonics intensity of each subsystem, their polarization angles deviate from that of the incoming laser pulse.
Starting from harmonic 9, we observe a different behavior. As $\alpha$ is increased, the polarization of the harmonics rotates in the opposite direction than the polarization of the driving field. This is caused by the phase difference of $\pi$ between the harmonics emitted by the two different TLS, effectively mirroring the polarization vector about the $x$-axis, which leads to the  'counter-rotation'  of the harmonic's polarization with respect to the polarization of the incoming laser field. \\

While this behavior looks similar to the two-center interference reported in HHG from aligned diatomic molecules \cite{lein2002role}, the microscopic origin is distinctly different. In H$_2^+$, interference between spatially separated emission centers produces similar amplitude minima and phase jumps for the high harmonic orders. In our model, in contrast, the phase behavior arises from two uncoupled TLS with distinct transition frequencies, $\omega_{21x}$ and $\omega_{30y}$, whose orthogonal dipoles lead to the observed signature in the amplitude and phase of the BTH.\\
\indent We expect similar features for the bound-bound contributions to BTH in aligned molecules with orthogonal transition dipole moments of different electronic transitions. Crystalline pentacene, recently shown to generate high-order harmonics with remarkable efficiency~\cite{wiechmann2025}, meets this criterion and presents a highly promising platform thanks to their intrinsic molecular alignment. Nevertheless, the continuum contributions to the generation of BTH in molecules might mask the observed features.\\
\indent In future work, our improved understanding of anisotropic phase features will advance existing HHG-based orbital tomography in aligned molecular targets. Focusing on the BTH may enable more detailed reconstructions sensitive to excited-state dynamics.


\begin{acknowledgments}
We acknowledge funding from the Deutsche Forschungsgemeinschaft via SFB 1477 'Light–Matter Interactions at Interfaces' (project no. 441234705).
\end{acknowledgments}


\bibliography{bib_new}

\widetext
\clearpage
\begin{center}
\textbf{\large Supplemental Material: Analytical derivation for the phase of the below-threshold harmonics in a two-level model system}
\end{center}
\setcounter{equation}{0}
\setcounter{figure}{0}
\setcounter{table}{0}
\setcounter{page}{1}
\makeatletter
\renewcommand{\theequation}{S\arabic{equation}}
\renewcommand{\thefigure}{S\arabic{figure}}

The succeeding derivation follows the manuscript of L. Plaja and L. Roso \cite{Plaja1993} very closely. Our derivation slightly differs in the treatment of the electric field and the phase after (\ref{eq_E}).\\

We start with the Hamiltonian 
\begin{equation}
    \hat{H}_0 = \begin{pmatrix}
0 & 0\\
0 & \omega_{10}
\end{pmatrix}, 
\end{equation}
that is the same as in our letter, just with shifted energies.
The interaction part is 
\begin{equation}
 \hat{H}_1(t) = \begin{pmatrix}
0 & E (t) d\\
E (t) d & 0
\end{pmatrix},
\end{equation}
with the electric field $E(t)$ and the transition dipole matrix element $d = \bra{\Psi_0 (0)} x \ket{\Psi_1 (0)}$.
We express the wavefunction 
\begin{equation}
     \ket{\Psi (t)} = C_0(t) \ket{\Psi_0 (0)} + C_1(t) \ket{\Psi_1 (0)}
\end{equation}
using time-dependent populations $C_0(t)$ and $C_1(t)$.
The time-dependent Schrödinger equation for the populations follows with
\begin{equation}
     \frac{d}{dt} C_0(t) = - i E(t) d C_1(t) \label{eq_c0}
\end{equation}
\begin{equation}
     \frac{d}{dt} C_1(t) = -i \omega_{10} C_1(t) - i E(t) d C_0(t).\label{eq_c1}
\end{equation}
From the populations we can calculate the dipole
\begin{align}
D(t) &= \bra{\Psi (t)} x \ket{\Psi (t)}   \nonumber\\&=d (C_0^*(t) C_1(t) + C_1^*(t) C_0(t)).
\end{align}
For the dipole velocity follows 
\begin{align}
\frac{d}{dt}D(t) = d [&\dot{C}_0^*(t) C_1(t) + C_0^*(t) \dot{C}_1(t) \nonumber\\+ &\dot{C}_1^*(t) C_0(t) + C_1^*(t) \dot{C}_0(t)],
\end{align}
and by plugging in (\ref{eq_c0}) and (\ref{eq_c1})
\begin{align}
\frac{d}{dt}D(t) &= d[
i E(t) d C_1^*(t)C_1(t) \nonumber
\\&\quad+C_0^*(t) (-i \omega_{10} C_1(t) - i E(t) d C_0(t))\nonumber
\\&\quad+(i \omega_{10} C_1^*(t) + i E(t) d C_0^*(t))C_0(t)\nonumber
\\&\quad-C_1^*(t) i E(t) d C_1(t)~]\nonumber\\
 &= - i \omega_{10} d [C_0^*(t)C_1(t) - C_1^*(t) C_0(t)]. \label{eq_dvel}
\end{align}
For the dipole acceleration follows
\begin{align}
\frac{d^2}{dt^2}D(t) &= - i \omega_{10} d [\dot{C}_0^*(t)C_1(t) +C_0^*(t)\dot{C}_1(t) \nonumber\\&\qquad- \dot{C}_1^*(t) C_0(t) - C_1^*(t) \dot{C}_0(t) ]\nonumber\\
&= - i \omega_{10} d [i E(t) d C^*_1(t)C_1(t) \nonumber\\
&\qquad+C_0^*(t)(-i \omega_{10} C_1(t) - i E(t) d C_0(t)) \nonumber\\
&\qquad-(i \omega_{10} C^*_1(t) + i E(t) d C^*_0(t)) C_0(t) \nonumber\\
&\qquad+C_1^*(t) i E(t) d C_1(t) ]\nonumber\\
&=-\omega_{10}^2\overbrace{d[C_0^*(t) C_1(t) + C^*_1(t)C_0(t)]}^{D(t)} \nonumber\\
&\qquad+2\omega_{10} E(t) d^2 [C^*_1(t)C_1(t) -C_0^*(t)C_0(t)],
\end{align}
and with the introduction of the population inversion $w(t) = |C_1(t)|^2 - |C_0(t)|^2$ 
\begin{equation}
    \frac{d^2}{dt^2}D(t) + \omega_{10}^2 D(t) = 2\omega_{10} E(t) d^2 w(t).\label{eq_dacc}
\end{equation}
Using (\ref{eq_c0}) and (\ref{eq_c1}) we can also derive an expression for the time-derivative of the population inversion by
\begin{align}
    \frac{d}{dt} w(t) &= \frac{d}{dt} [C^*_1(t) C_1(t) - C^*_0(t) C_0(t)]\nonumber\\
    &= \dot{C}^*_1(t) C_1(t) + C^*_1(t) \dot{C}_1(t) \nonumber\\
    &\quad- \dot{C}^*_0(t) C_0(t) -  C^*_0(t) \dot{C}_0(t)\nonumber\\
    &= [i \omega_{10} C_1(t)^* + i E(t) d C_0^*(t)] C_1(t) \nonumber\\
    &\quad+ C^*_1(t) [-i \omega_{10} C_1(t) - i E(t) d C_0(t)] \nonumber\\
    &\quad-  i E(t) d C_1^*(t) C_0(t) -  C^*_0(t) [- i E(t) d C_1(t)]\nonumber\\
    &=2 i E(t) d \left[C^*_0(t) C_1(t) - C^*_1(t) C_0(t)\right].
\end{align}
With (\ref{eq_dvel}) follows
\begin{equation}
     \frac{d}{dt} w(t) = - 2 \frac{E(t)}{\omega_{10}} \frac{d}{dt}D(t).\label{eq_w}
\end{equation}\\

We consider a periodic electric field 
\begin{equation}
     E(t) = E_0 \cos(\omega_L t - \varphi) \label{eq_E}.
\end{equation}
If our system is only populating a single Floquet state of the interacting system, all dynamics are adiabatic, and the dipole and the population inversion will be periodic with the laser period \cite{Bauer2013}. Therefore, they can be expressed in terms of multiples of the laser frequency by the Fourier expansion 
\begin{align}
     w(t) &= \sum_n w_n e^{i n \omega_L t}\nonumber\\
     D(t) &= \sum_n d_n e^{i n\omega_L t},\label{eq_fourier}
\end{align}\\
with the time-independent coefficients $w_n$ and $d_n$, and the index $n$ running over all integer numbers.
Plugging (\ref{eq_fourier}) into (\ref{eq_w})
yields
\begin{align}
    \frac{d}{dt} \sum_n w_n e^{i n \omega_L t} &= \sum_n -2 \frac{E_0}{\omega_{10}} \cos(\omega_L t - \varphi) \frac{d}{dt} d_n e^{in\omega_L t}\nonumber\\
    \sum_n n w_n e^{i n \omega_L t} &=
    \sum_n -\frac{E_0}{\omega_{10}} n d_n e^{in\omega_L t}(e^{i\varphi} e^{-i\omega_Lt} + e^{-i\varphi}e^{i\omega_L t})\nonumber\\
    &=\sum_n -\frac{E_0}{\omega_{10}} n d_n (e^{i\varphi} e^{i(n-1)\omega_Lt} + e^{-i\varphi}e^{i(n+1)\omega_L t})\nonumber\\
    &=\sum_n -\frac{E_0}{\omega_{10}} \left[(n+1)d_{n+1}e^{i\varphi} + (n-1)d_{n-1} e^{-i\varphi}\right]e^{in\omega_L t},
\end{align}
and by comparing the coefficients we obtain
\begin{equation}
    n w_n =  -\frac{E_0}{\omega_{10}} \left[(n+1) d_{n+1}e^{i\varphi} + (n-1) d_{n-1} e^{-i\varphi}\right].\label{eq_wq}
\end{equation}
Analogously, by inserting the Fourier expansions~(\ref{eq_fourier}) into (\ref{eq_dacc}), 
\begin{align}
\sum_n \left[(i n w_L)^2 d_n e^{in\omega_L t} + \omega_{10}^2  d_n e^{in\omega_L t} \right]
&=\sum_n 2\omega_{10}  E_0 \cos(\omega_L t - \varphi) d^2 w_n e^{i n \omega_L t}\nonumber\\
&=\sum_n \omega_{10} E_0 d^2 w_n e^{i n \omega_L t} (e^{i\varphi} e^{-i\omega_Lt} + e^{-i\varphi}e^{i\omega_L t})\nonumber\\
&=\sum_n \omega_{10} E_0 d^2 (w_{n+1}e^{i\varphi} + w_{n-1}e^{-i\varphi})e^{i n \omega_L t},
\end{align}
and again comparing the coefficients,
\begin{equation}
   (\omega_{10}^2 - n^2 w_L^2)  d_n = \omega_{10} E_0 d^2 (w_{n+1}e^{i\varphi} + w_{n-1}e^{-i\varphi})\label{eq_wq2}
\end{equation}
follows.
We can utilize (\ref{eq_wq}) to write
\begin{align}
w_{n+1} &=  -\frac{1}{n+1}\frac{E_0}{\omega_{10}} [(n+2) d_{n+2}e^{i\varphi} + n d_{n} e^{-i\varphi}], \quad\text{and} \\
w_{n-1} &=  -\frac{1}{n-1}\frac{E_0}{\omega_{10}} [n d_{n}e^{i\varphi} + (n-2) d_{n-2} e^{-i\varphi}],
\end{align}
and insert these expressions into (\ref{eq_wq2}), to obtain a continued-fractions expression for the dipole coefficients 
\begin{align}
(\omega_{10}^2 - n^2 w_L^2)  d_n &=
-\omega_{10} E_0 d^2 \left(\frac{1}{n+1}\frac{E_0}{\omega_{10}} [(n+2) d_{n+2}e^{i\varphi} + n d_{n} e^{-i\varphi}]e^{i\varphi} \right.\nonumber\\ &\quad+ \left.\frac{1}{n-1}\frac{E_0}{\omega_{10}} [n d_{n}e^{i\varphi} + (n-2) d_{n-2} e^{-i\varphi}]e^{-i\varphi}\right)\nonumber\\
\frac{\omega_{10}^2 - n^2 w_L^2}{E_0^2 d^2}  d_n&=-\left[\frac{n+2}{n+1} d_{n+2}e^{2i\varphi} + \left(\frac{n}{n+1} + \frac{n}{n-1}\right) d_{n}  + \frac{n-2}{n-1} d_{n-2} e^{-2i\varphi}\right]\nonumber\\
\left(\frac{2 n^2}{n^2 -1} + \frac{\omega_{10}^2 - n^2 w_L^2}{E_0^2 d^2} \right) d_n&=-\left(\frac{n+2}{n+1} d_{n+2}e^{2i\varphi} + \frac{n-2}{n-1} d_{n-2} e^{-2i\varphi}\right)\nonumber\\
    d_n &= -\left(\frac{n+2}{n+1} d_{n+2}e^{2i\varphi} + \frac{n-2}{n-1} d_{n-2} e^{-2i\varphi}\right) \left(\frac{2 n^2}{n^2 -1} + \frac{\omega_{10}^2 - n^2 w_L^2}{E_0^2 d^2} \right) ^{-1}. \label{eq_confrac}
\end{align}
For a parameter regime, where each successive harmonic order is substantially weaker than the previous one, we can neglect the $d_{n+2}$-term in (\ref{eq_confrac}), resulting in the expression
\begin{equation}
    d_n = -e^{-2i\varphi}\frac{n-2}{n-1} d_{n-2} \left(\frac{2 n^2}{n^2 -1} + \frac{\omega_{10}^2 - n^2 w_L^2}{E_0^2 d^2} \right) ^{-1}. 
\end{equation}
Due to the nature of the Fourier expansion, the complex coefficient $d_n=\widetilde{D}(n\omega_L)$ is equivalent to the Fourier transform of the dipole at the frequency $n\omega_L$. Therefore, the phase of the $n$-th harmonic can be extracted by $\varphi_n = \text{arg}(d_n)$.

\end{document}